\documentclass[12pt]{elsarticle}
\usepackage{amssymb}
\usepackage{amsmath}%[fleqn]
\usepackage{amsthm}
\usepackage{subfigure}
\usepackage{graphicx}

\biboptions{numbers,sort&compress}
\usepackage[dvipdfm,colorlinks,linkcolor=red,anchorcolor=blue,citecolor=blue]{hyperref}
\usepackage{booktabs}
\makeatletter

\newcommand{\Rmnum}[1]{\expandafter\@slowromancap\romannumeral #1@}
\makeatother
\usepackage{caption}

\usepackage{anysize}
\usepackage{setspace}
\marginsize{2.0cm}{1.5cm}{0.6cm}{0.6cm} %%LRTB
\journal{Journal of XX}

\begin{document}

\begin{frontmatter}

\title{Periodic and rational solutions of the reduced Maxwell-Bloch equations }
\author{Jiao Wei$^{{\rm a}}$}%\ead{weijiaozzu@163.com}
\author{Xin Wang$^{{\rm a,b}}$}\ead{wangxinlinzhou@163.com}
\author{Xianguo Geng$^{{\rm a}}$}\ead{xggeng@zzu.edu.cn}

\cortext[cor1]{Corresponding author.}
\address{$^{{\rm a}}$School of Mathematics and Statistics, Zhengzhou University, 100 Kexue Road,
Zhengzhou, Henan, 450001,  China}
\address{$^{{\rm b}}$College of Science, Zhongyuan University of Technology, Zhengzhou, 450007, China}

\begin{abstract}
We investigate the reduced Maxwell-Bloch (RMB) equations which describe the propagation of short optical pulses
in dielectric materials with resonant non-degenerate transitions.
The general $N$th-order periodic solutions are provided by means of the Darboux transformation,
and from two different limiting cases of the obtained general periodic solutions,
the $N$th-order degenerate periodic and $N$th-order rational solutions containing several free parameters
with compact determinant representations are derived, respectively.
Explicit expressions of these solutions from first to second order are presented.
Typical nonlinear wave patterns for the four components of the RMB equations
such as single-peak, double-peak-double-dip, double-peak and single-dip structures
in the second-order rational solutions are shown.
This kind of the rational solutions correspond to rogue waves in the reduced Maxwell-Bloch equations. 
\end{abstract}

\end{frontmatter}

\section{Introduction }  %%% ½Ú±êÌâ 2

The associated reduced Maxwell-Bloch (RMB) equations play a fundamental role to describe wave phenomena
in nonlinear optics related to self-induced transparency \cite{01}. In dimensionless form, as
\begin{subequations}\label{01}
\begin{align}
&u_{x}=-\mu v,\label{1a}\\
&v_{x}=Ew+\mu u,\label{1b}\\
&w_{x}=-Ev,\label{1c}\\
&E_{t}=-v,\label{1d}
\end{align}
\end{subequations}
with $E(x,t)$ the electric field, $u(x,t)$ atomic dipole, $v(x,t)$ phase information and
$w(x,t)$ the atomic inversion.
The integrability such as the Painlev\'{e} test and Lie-algebra-valued differential
forms of the RMB equations have been investigated in Refs. \cite{02,03}, and the
explicit $N$-soliton solutions of the RMB equations have been respectively studied by the
inverse scattering transform, Hirota bilinear technique and Darboux transformation (DT)
during the past few decades \cite{04,05,06,07,08}.

Recently, it is well known that the generation of unexpectedly huge waves (termed as \lq\lq rogue waves\rq\rq)
has received widespread attention in quite a lot of researches including
oceanography, optical fields, Bose-Einstein condensates, plasma physics, etc. \cite{09,10,11,12}.
The straightforward description of a single rogue wave in mathematics is
the Peregrine soliton \cite{13}, a special solution of the nonlinear Schr\"{o}dinger (NLS) equation,
which is a combination of the second-order rational polynomials and exponential function,
and simulates the evolution of a wave of large amplitude that is localized in both space and time.
More recently, beyond the NLS equation and its relevant physical systems  \cite{14,15,16,17,18,19,20,21},
explicit periodic solutions, rational solutions and the generation of
rogue waves in the modified
Korteweg-de Vries (mKdV) equation have been studied by Chowdury, Slunyaev and He
et al. \cite{22,23,He}.
As is pointed out by them, the existences of periodic and rational solutions
in the mKdV equation reveal that breather and rogue wave phenomena are not confined to the
deep ocean, and rogue wave phenomena governed by the mKdV equation present
quite different descriptions in hydrodynamics from that related to the NLS equation.

In this paper, we demonstrate that Eq. (\ref{01}) can also possess periodic and rational
solutions like the mKdV equation, which will be helpful to understand the complicated rogue wave
phenomena in nonlinear optics governed by the RMB equations. We present the general $N$th-order periodic
solutions on a finite constant background by using the classical DT with $N$ eigenvalues that are different
from each other \cite{24,25,26,27}.
Then, by taking advantage of the limit approach, namely the generalized DT \cite{28,29,30,31,32,WL1,WL2,WL3},
the $N$th-order degenerate periodic and  $N$th-order rational solutions in the compact determinant representations can be respectively derived from two kinds of limiting cases of the general periodic solutions.
As an application, explicit periodic, degenerate periodic and rational solutions up to second order are presented.
We hereby show that the doubly-periodic lattice-like structure, and the single periodic- peaks or dips on
a periodic wave background structure can exist in the second-order periodic and degenerate periodic solutions, respectively. Particularly, we demonstrate that, the second-order rational solutions for the four
components $E$, $v$, $w$ and $u$ in Eq. (\ref{01}) can provide distinctive patterns as a result of the collisions of a fixed number of dark and bright solitons, namely, the single-peak, double-peak-double-dip, double-peak and single-dip structures, respectively.
Further, it is  computed that the maximum amplitudes of the rational solutions
from first to fourth order for the electric field $E$
are the same as that of rogue waves from first to fourth order for the NLS equation \cite{28}. Finally, it is confirmed that the free parameters can produce an important \lq\lq differential shift\rq\rq \cite{22}
effect on the peaks or dips with maximum or minimum amplitudes in the rational solutions.

The present paper is constructed as follows. In section 2, the general
$N$th-order periodic solutions are given by utilizing the classical DT.
In sections 3 and 4, the $N$th-order rational and $N$th-order degenerate periodic solutions
are derived through two different limit approaches, respectively.
Explicit expressions of these obtained solutions from first to second order are presented,
and some interesting wave patterns are shown. The last section is the conclusion of this paper.

\section{Periodic solutions}

For our studies, we begin with the Lax pair of Eq. (\ref{01}) which can be given through the
Ablowitz-Kaup-Newell-Segur (AKNS) technique \cite{AKNS,G1,G2,G3}:
\begin{align}
&\Phi_{x}=U\Phi,U=
\left(
  \begin{array}{cc}
    -i\eta & -\dfrac{1}{2}E \\
   \dfrac{1}{2}E & i\eta \\
  \end{array}
\right),\label{02}\\
&\Phi_{t}=V\Phi,V=\dfrac{1}{4\eta^2-\mu^2}
\left(
  \begin{array}{cc}
    -i\eta w & -i\eta v+\dfrac{1}{2}\mu u \\
    -i\eta v-\dfrac{1}{2}\mu u & i\eta w \\
  \end{array}
\right).\label{03}
\end{align}
The compatibility condition $U_{t}-V_{x}+UV-VU=0$ can directly give rise to Eq. (\ref{01}).

Next, in order to obtain periodic solutions from the classical DT, we choose the
following constant seeding solutions
\begin{equation}\label{04}
E[0]=e_{0},v[0]=0,w[0]=-\dfrac{\mu u_{0}}{e_{0}},u[0]=u_{0}.
\end{equation}
By substituting the above solution into the linear system (\ref{02}) and (\ref{03}), and setting the eigenvalue
$\eta=ie_{0}(1/2+\epsilon^2)$ with $\epsilon$ being the pure imaginary small parameter such that
$|\epsilon|<1$, we have the following solution
\begin{equation}\label{05}
\Phi(\epsilon)=\left(
       \begin{array}{c}
         \psi \\
         \varphi \\
       \end{array}
     \right)=
\left(
  \begin{array}{c}
    C_{1}e^{A}-C_{2}e^{-A} \\
    C_{2}e^{A}-C_{1}e^{-A} \\
  \end{array}
\right),
\end{equation}
where
$$
C_{1}=\frac{(1+2\epsilon^2+2\epsilon\sqrt{1+\epsilon^2})^{\frac{1}{2}}}{2\epsilon\sqrt{1+\epsilon^2}},
C_{2}=\frac{(1+2\epsilon^2-2\epsilon\sqrt{1+\epsilon^2})^{\frac{1}{2}}}{2\epsilon\sqrt{1+\epsilon^2}},
$$
and
$$
A=e_{0}\epsilon\sqrt{1+\epsilon^2}\left[x+\dfrac{\mu u_{0}}{e_{0}(4e_{0}^2\epsilon^4
+4e_{0}^2\epsilon^2+e_{0}^2+\mu^2)}t   \right].
$$
After that, by letting $\Phi_{j}=(\psi_{j},\varphi_{j})^{T}=\Phi(\epsilon)|_{\epsilon=\epsilon_{j}}$ ($j=1,2,\cdots,N$) be $N$ special solutions of the linear system (\ref{02}) and (\ref{03}) with the constant seeding solutions (\ref{04}) and
$\eta_{j}=ie_{0}(1/2+\epsilon_{j}^2)$, here $\epsilon_{i}\neq\epsilon_{j}$ for $i\neq j$. Thus,
we can obtain the general $N$th-order periodic solutions via the classical DT, viz.
\begin{align}
&E[N]=e_{0}-4i\dfrac{\det(M_{1})}{\det(M)},\label{06}\\
&v[N]=4i\dfrac{\partial}{\partial t}\dfrac{\det(M_{1})}{\det(M)},\label{07}\\
&w[N]=-\dfrac{\mu u_{0}}{e_{0}}+4i\dfrac{\partial}{\partial t}\dfrac{\det(M_{2})}{\det(M)},\label{08}
\end{align}
and
\begin{equation}\label{09}
u[N]=u_{0}+\dfrac{4i}{\mu}\left[\dfrac{\partial^2}{\partial x\partial t}\dfrac{\det(M_{1})}{\det(M)}-
e_{0}\dfrac{\partial }{\partial t}\dfrac{\det(M_{2})}{\det(M)}
-\dfrac{\mu u_{0}}{e_{0}}\dfrac{\det(M_{1})}{\det(M)}+4i\dfrac{\det(M_{1})}{\det(M)}\dfrac{\partial }{\partial t}\dfrac{\det(M_{2})}{\det(M)}\right],
\end{equation}
where
$$\begin{array}{l}
M=\left(
    \begin{array}{cccc}
      M_{11} & M_{12} & \cdots & M_{1N} \\
       M_{21} & M_{22} & \cdots & M_{2N} \\
      \vdots & \vdots & \ddots & \vdots \\
       M_{N1} & M_{N2} & \cdots & M_{NN} \\
    \end{array}
  \right),\\
M_{1}=\left(
    \begin{array}{ccccc}
      M_{11} & M_{12} & \cdots & M_{1N} & \varphi_{1}\\
       M_{21} & M_{22} & \cdots & M_{2N} & \varphi_{2}\\
      \vdots & \vdots & \ddots & \vdots \vdots\\
       M_{N1} & M_{N2} & \cdots & M_{NN} & \varphi_{N}\\
       \psi_{1} &\psi_{2} &\cdots &\psi_{N}& 0
    \end{array}
  \right),\end{array}$$

$$\begin{array}{l}
M_{2}=\left(
    \begin{array}{ccccc}
      M_{11} & M_{12} & \cdots & M_{1N} & \psi_{1}\\
       M_{21} & M_{22} & \cdots & M_{2N} & \psi_{2}\\
      \vdots & \vdots & \ddots & \vdots \vdots\\
       M_{N1} & M_{N2} & \cdots & M_{NN} & \psi_{N}\\
       \psi_{1} &\psi_{2} &\cdots &\psi_{N}& 0
    \end{array}
  \right),
\end{array}$$
with
$$
M_{ij}=\dfrac{\psi_{j}\psi_{i}+\varphi_{j}\varphi_{i}}{ie_{0}(1+\epsilon_{j}^2+\epsilon_{i}^2)},1\leq i,j\leq N.
$$

At this point, by following the formulas (\ref{06})-(\ref{09}) with $N=1$, we can present the first-order periodic solutions
\begin{align}
&E[1]_{p}=e_{0}\left[1-2\dfrac{(2\kappa_{1}^2-1)+\cos(2\rho_{1})}{1/(2\kappa_{1}^2-1)
+\cos(2\rho_{1})}\right],\label{10}\\
&v[1]_{p}=-2e_{0}\dfrac{\partial}{\partial t}\left[\dfrac{(2\kappa_{1}^2-1)+\cos(2\rho_{1})}{1/(2\kappa_{1}^2-1)+\cos(2\rho_{1})}\right],\label{11}\\
&w[1]_{p}=-\dfrac{\mu u_{0}}{e_{0}}+2e_{0}\dfrac{\partial}{\partial
t}\left[\dfrac{2\kappa_{1}\sqrt{1-\kappa_{1}^2}\sin(2\rho_{1})
+(2\kappa_{1}^2-1)\cos(2\rho_{1})+1}{1/(2\kappa_{1}^2-1)+\cos(2\rho_{1})}\right],\label{12}
\end{align}
and
\begin{equation}\label{13}
u[1]_{p}=u_{0}+\dfrac{4i}{\mu}\left[\dfrac{\partial^2}{\partial x\partial t}\dfrac{G_{p}^{[1]}}{F_{p}^{[1]}}-
e_{0}\dfrac{\partial }{\partial t}\dfrac{H_{p}^{[1]}}{F_{p}^{[1]}}-\dfrac{\mu u_{0}}{e_{0}}\dfrac{G_{p}^{[1]}}{F_{p}^{[1]}}
+4i\dfrac{G_{p}^{[1]}}{F_{p}^{[1]}}\dfrac{\partial }{\partial t}\dfrac{H_{p}^{[1]}}{F_{p}^{[1]}}\right],
\end{equation}
where $\epsilon_{1}=i\kappa_{1}$ with $\kappa_{1}$
being a real small parameter and satisfying $|\kappa_{1}|<1$,
$$
\begin{array}{l}
\rho_{1}=e_{0}\kappa_{1}\sqrt{1-\kappa_{1}^2}\left[x+\dfrac{\mu u_{0}}{e_{0}(4e_{0}^2\kappa_{1}^4
-4e_{0}^2\kappa_{1}^2+e_{0}^2+\mu^2)}t\right],\\
F_{p}^{[1]}=\dfrac{2\kappa_{1}^2\cos(2\rho_{1})-\cos(2\rho_{1})+1}{ie_{0}\kappa_{1}^2(2\kappa_{1}^2-1)
(\kappa_{1}^2-1)},
G_{p}^{[1]}=-\dfrac{1}{2}\dfrac{[2\kappa_{1}^2+\cos(2\rho_{1})-1]}{\kappa_{1}^2(\kappa_{1}^2-1)},\\
H_{p}^{[1]}=\dfrac{1}{2}\dfrac{[2\kappa_{1}\sqrt{1-\kappa_{1}^2}\sin(2\rho_{1})+2\kappa_{1}^2\cos(2\rho_{1})
-\cos(2\rho_{1})+1]}{\kappa_{1}^2(\kappa_{1}^2-1)}.
\end{array}
$$

\begin{figure}[!h]
\centering
\renewcommand{\figurename}{{\bf Fig.}}
{\includegraphics[height=10cm,width=16cm]{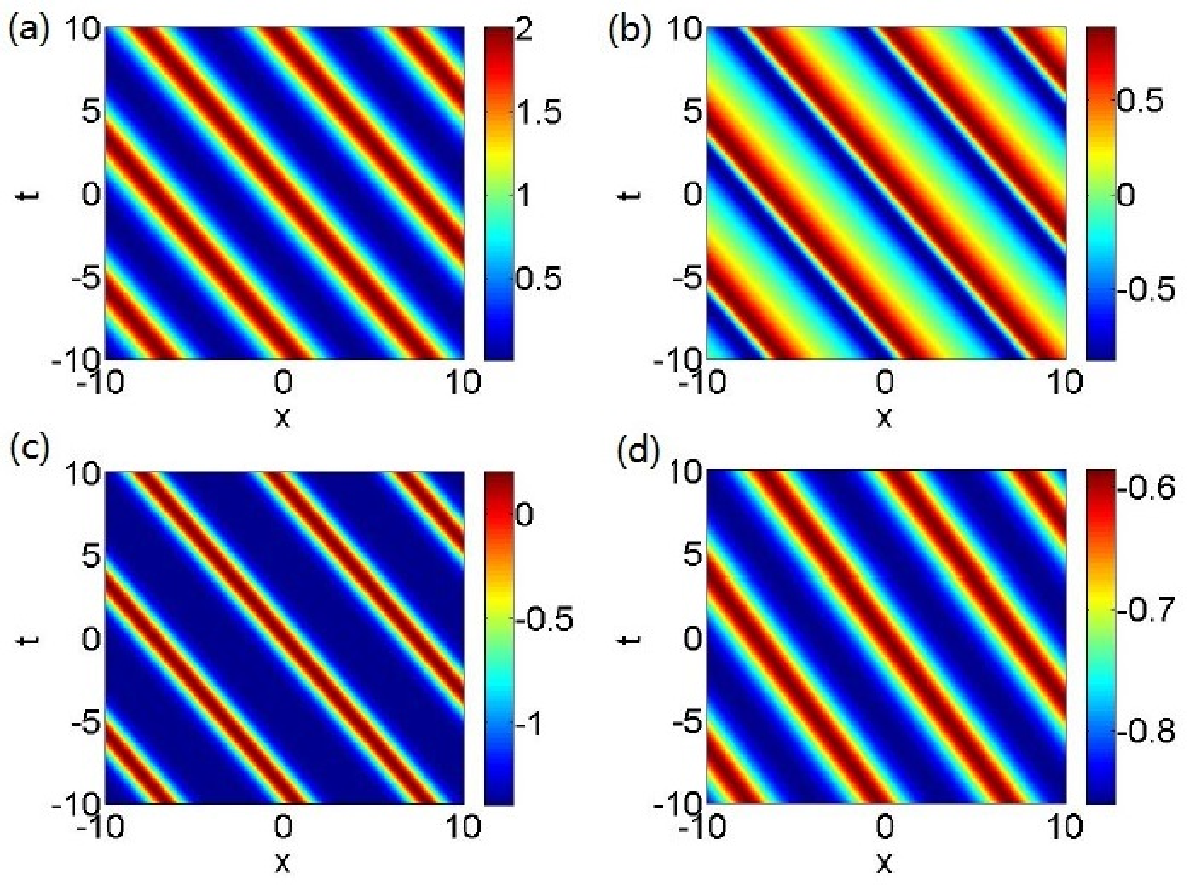}}
\caption{(a)-(d) The first-order periodic solutions (\ref{10})-(\ref{12}) for the components $E$, $v$, $w$
with $\mu=1$ and (\ref{13}) for the component $u$ with $\mu=0.2$. The other parameters are $e_{0}=1,u_{0}=1,\kappa_{1}=0.5$. }
\label{fig:1}
\end{figure}

The first-order periodic solutions (\ref{10})-(\ref{13}) are shown in Fig. \ref{fig:1}.
It is seen that these solutions are periodic in both $x$ and $t$
and maintain constant amplitudes. The maximum amplitudes of the components
$E$, $v$, $w$ and $u$ are 2, 0.88, 0.20  and -0.59, respectively.

For $N=2$ in the formulas (\ref{06})-(\ref{09}), the second-order periodic solutions can be worked out as
\begin{align}
&E[2]_{p}=e_{0}-4i\dfrac{G_{p}^{[2]}}{F_{p}^{[2]}},\label{14}\\
&v[2]_{p}=4i\dfrac{\partial}{\partial t}\dfrac{G_{p}^{[2]}}{F_{p}^{[2]}},\label{15}\\
&w[2]_{p}=-\dfrac{\mu u_{0}}{e_{0}}+4i\dfrac{\partial}{\partial t}\dfrac{H_{p}^{[2]}}{F_{p}^{[2]}},\label{16}
\end{align}
and
\begin{equation}\label{17}
u[2]_{p}=u_{0}+\dfrac{4i}{\mu}\left(\dfrac{\partial^2}{\partial x\partial t}
\dfrac{G_{p}^{[2]}}{F_{p}^{[2]}}-
e_{0}\dfrac{\partial }{\partial t}\dfrac{H_{p}^{[2]}}{F_{p}^{[2]}}-\dfrac{\mu u_{0}}{e_{0}}\dfrac{G_{p}^{[2]}}{F_{p}^{[2]}}
+4i\dfrac{G_{p}^{[2]}}{F_{p}^{[2]}}\dfrac{\partial }
{\partial t}\dfrac{H_{p}^{[2]}}{F_{p}^{[2]}}\right),
\end{equation}
where
$$
\begin{array}{l}
F_{p}^{[2]}=F_{11}F_{22}-F_{12}F_{21},
G_{p}^{[2]}=-F_{11}c_{2}d_{2}+F_{12}c_{1}d_{2}+F_{21}c_{2}d_{1}-F_{22} c_{1} d_{1},\\
H_{p}^{[2]}=-F_{11}c_{2}^2+F_{12}c_{1}c_{2}+F_{21}c_{1}c_{2}-F_{22}c_{1}^2,
\end{array}
$$
with
$$
\begin{array}{l}
F_{11}=\dfrac{2\kappa_{1}^2\cos(2\rho_{1})-\cos(2\rho_{1})+1}{ie_{0}\kappa_{1}^2(2\kappa_{1}^2-1)
(\kappa_{1}^2-1)},F_{22}=\dfrac{2\kappa_{2}^2\cos(2\rho_{2})-\cos(2\rho_{2})+1}
{ie_{0}\kappa_{2}^2(2\kappa_{2}^2-1)
(\kappa_{2}^2-1)},\\
F_{12}=F_{21}=\dfrac{2i[\sqrt{1-\kappa_{1}^2}\sqrt{1-\kappa_{2}^2}\sin(\rho_{1})\sin(\rho_{2})
+\kappa_{1}\kappa_{2}\cos(\rho_{1})\cos(\rho_{2})]}
{e_{0}\kappa_{1}\kappa_{2}(\kappa_{1}^2+\kappa_{2}^2-1)\sqrt{1-\kappa_{1}^2}\sqrt{1-\kappa_{2}^2}},\\
c_{1}=\dfrac{\sqrt{1-\kappa_{1}^2}\sin(\rho_{1})
+\kappa_{1}\cos(\rho_{1})}{\kappa_{1}\sqrt{1-\kappa_{1}^2}},c_{2}=
\dfrac{\sqrt{1-\kappa_{2}^2}\sin(\rho_{2})+\kappa_{2}\cos(\rho_{2})}{\kappa_{2}\sqrt{1-\kappa_{2}^2}},\\
d_{1}=\dfrac{\sqrt{1-\kappa_{1}^2}\sin(\rho_{1})
-\kappa_{1}\cos(\rho_{1})}{\kappa_{1}\sqrt{1-\kappa_{1}^2}},d_{2}=
\dfrac{\sqrt{1-\kappa_{2}^2}\sin(\rho_{2})-\kappa_{2}\cos(\rho_{2})}{\kappa_{2}\sqrt{1-\kappa_{2}^2}},\\
\rho_{j}=e_{0}\kappa_{j}\sqrt{1-\kappa_{j}^2}\left[x+\dfrac{\mu u_{0}}{e_{0}(4e_{0}^2\kappa_{j}^4
-4e_{0}^2\kappa_{j}^2+e_{0}^2+\mu^2)}t\right],j=1,2.
\end{array}
$$
Here we take $\epsilon_{j}=i\kappa_{j}$ with $\kappa_{j}$ being a real parameter and
yielding $|\kappa_{j}|<1$.

\begin{figure}[!h]
\centering
\renewcommand{\figurename}{{\bf Fig.}}
{\includegraphics[height=10cm,width=16cm]{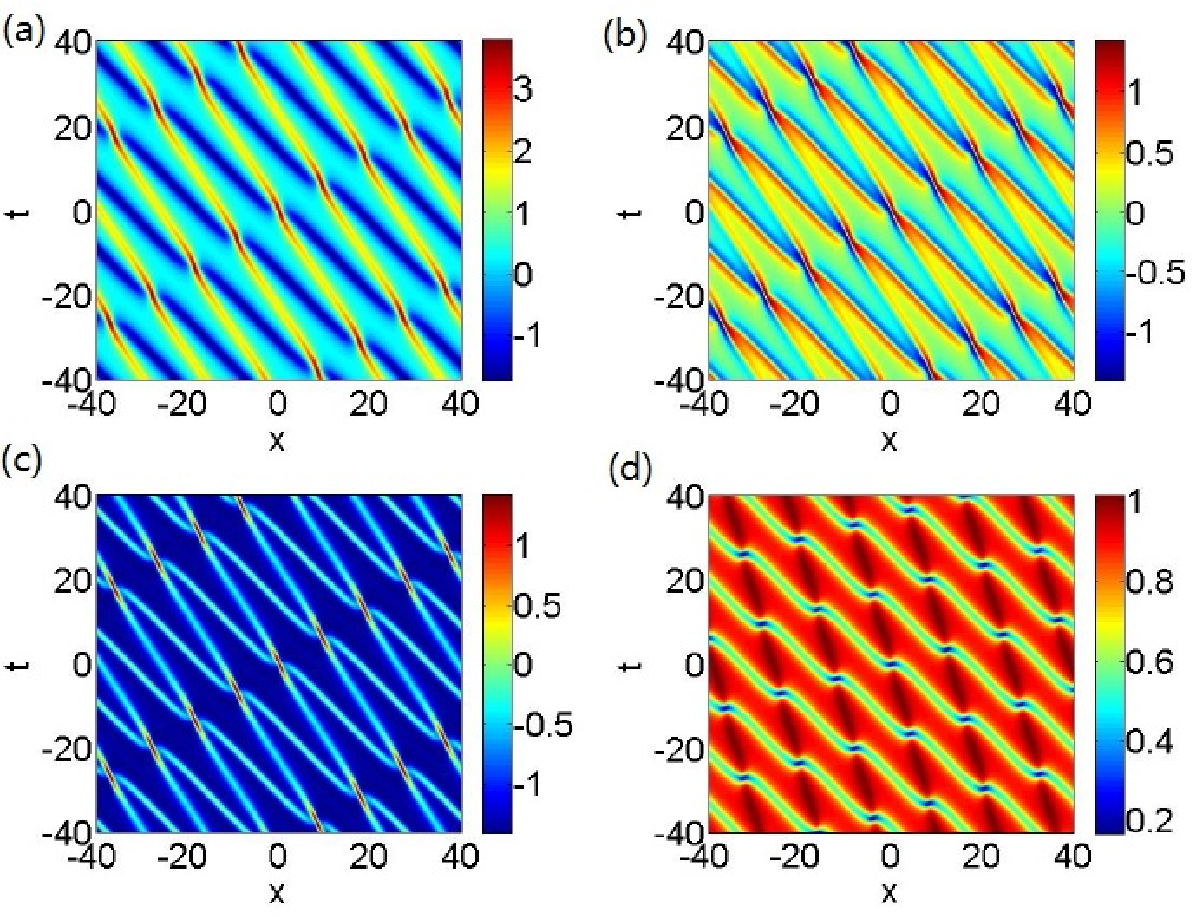}}
\caption{(a)-(d) The second-order periodic solutions (\ref{14})-(\ref{16}) for the components $E$, $v$, $w$
with $\mu=1$ and (\ref{17}) for the component $u$ with $\mu=0.2$. The other parameters are $e_{0}=1,u_{0}=1,\kappa_{1}=0.5,\kappa_{2}=0.25$.}
\label{fig:2}
\end{figure}

In this circumstance, the doubly-periodic lattice-like structures can be displayed, see Fig. \ref{fig:2}.
The maximum amplitudes of the peaks for the components $E$, $v$ and $w$ are 3.75, 1.41 and 1.41.
While for the component $u$, the doubly-periodic lattice-like dip structure appears, and the
minimum amplitude of the dips is 0.17.

\section{Degenerate periodic solutions}

In this section, we derive the degenerate periodic solutions of Eq. (\ref{01}) from one
limiting case of the general $N$th-order periodic solutions (\ref{06})-(\ref{09}).
To this end, we give the following  Taylor series
\begin{equation}\label{18}
\psi(\epsilon)=\sum_{i=0}^{N-1}\psi_{1}^{[i]}(\epsilon-\epsilon_{1})^{i}
+\mathcal{O}\left((\epsilon-\epsilon_{1})^N\right),
\varphi(\epsilon)=\sum_{i=0}^{N-1}\varphi_{1}^{[i]}(\epsilon-\epsilon_{1})^{i}
+\mathcal{O}\left((\epsilon-\epsilon_{1})^N\right),
\end{equation}
and define
\begin{equation}\label{19}
\dfrac{\psi(\epsilon^{*})\psi(\epsilon)+\varphi(\epsilon^{*})\varphi(\epsilon)}{ie_{0}(1+\epsilon^2+\epsilon^{*2})}
=\sum_{i,j=1}^{N}P^{[i,j]}(\epsilon-\epsilon_{1})^{j-1}(\epsilon^{*}-\epsilon_{1})^{i-1}+
\mathcal{O}\left((\epsilon-\epsilon_{1})^N(\epsilon^{*}-\epsilon_{1})^N\right),
\end{equation}
where
$$
\psi_{1}^{[i]}=\lim\limits_{\epsilon\rightarrow\epsilon_{1}}
\frac{\displaystyle1}{\displaystyle i!}\frac{\displaystyle\partial^{i}\psi_{1}}{\displaystyle\partial \epsilon^{i}},\varphi_{1}^{[i]}=\lim\limits_{\epsilon\rightarrow\epsilon_{1}}
\frac{\displaystyle1}{\displaystyle i!}\frac{\displaystyle\partial^{i}\varphi_{1}}{\displaystyle\partial \epsilon^{i}},
$$
and
$$
P^{[i,j]}=\dfrac{1}{(i-1)!(j-1)!}\dfrac{\partial ^{i+j-2}}{\partial \epsilon^{j-1}\partial \epsilon^{*i-1}}
\dfrac{\psi(\epsilon^{*})\psi(\epsilon)+\varphi(\epsilon^{*})\varphi(\epsilon)}{ie_{0}(1+\epsilon^2+\epsilon^{*2})}
\bigg|_{\epsilon,\epsilon^{*}\rightarrow\epsilon_{1}}.
$$
Here $\epsilon^{*}$ is the other introduced complex small parameter, and $\epsilon_{1}$ is
a pure imaginary small parameter such that $\epsilon_{1}\neq0$.
At this time, the $N$th-order degenerate periodic solutions can be expressed as
\begin{align}
&E[N]=e_{0}-4i\dfrac{\det(P_{1})}{\det(P)},\label{20}\\
&v[N]=4i\dfrac{\partial}{\partial t}\dfrac{\det(P_{1})}{\det(P)},\label{21}\\
&w[N]=-\dfrac{\mu u_{0}}{e_{0}}+4i\dfrac{\partial}{\partial t}\dfrac{\det(P_{2})}{\det(P)},\label{22}
\end{align}
and
\begin{equation}\label{23}
u[N]=u_{0}+\dfrac{4i}{\mu}\left[\dfrac{\partial^2}{\partial x\partial t}\dfrac{\det(P_{1})}{\det(P)}-
e_{0}\dfrac{\partial }{\partial t}\dfrac{\det(P_{2})}{\det(P)}
-\dfrac{\mu u_{0}}{e_{0}}\dfrac{\det(P_{1})}{\det(P)}+4i\dfrac{\det(P_{1})}{\det(P)}\dfrac{\partial }{\partial t}\dfrac{\det(P_{2})}{\det(P)}\right],
\end{equation}
where
$$\begin{array}{l}
P=\left(
    \begin{array}{cccc}
      P^{[11]} & P^{[12]} & \cdots & P^{[1N]} \\
       P^{[21]} & P^{[22]} & \cdots & P^{[2N]} \\
      \vdots & \vdots & \ddots & \vdots \\
       P^{[N1]} & P^{[N2]} & \cdots & P^{[NN]} \\
    \end{array}
  \right),\\
P_{1}=\left(
    \begin{array}{ccccc}
      P^{[11]} & P^{[12]} & \cdots & P^{[1N]} & \varphi_{1}^{[0]}\\
       P^{[21]} & P^{[22]} & \cdots & P^{[2N]}  & \varphi_{1}^{[1]}\\
      \vdots & \vdots & \ddots & \vdots \vdots\\
       P^{[N1]} & P^{[N2]} & \cdots & P^{[NN]} & \varphi_{1}^{[N-1]}\\
       \psi_{1}^{[0]} &\psi_{1}^{[1]} &\cdots &\psi_{1}^{[N-1]}& 0
    \end{array}
  \right),\\
P_{2}=\left(
    \begin{array}{ccccc}
      P^{[11]} & P^{[12]} & \cdots & P^{[1N]} & \psi_{1}^{[0]}\\
        P^{[21]} & P^{[22]} & \cdots & P^{[2N]}  & \psi_{1}^{[1]}\\
      \vdots & \vdots & \ddots & \vdots \vdots\\
       P^{[N1]} & P^{[N2]} & \cdots & P^{[NN]} & \psi_{1}^{[N-1]}\\
       \psi_{1}^{[0]} &\psi_{1}^{[1]} &\cdots &\psi_{1}^{[N-1]}& 0
    \end{array}
  \right).
\end{array}
$$

Explicitly, for simplicity, we now choose a special small parameter of $\epsilon_{1}=1/2i$,
then the degenerate periodic solutions take the form as
\begin{align}
&E[2]_{d}=e_{0}-4i\dfrac{G_{d}^{[2]}}{F_{d}^{[2]}},\label{24}\\
&v[2]_{d}=4i\dfrac{\partial}{\partial t}\dfrac{G_{d}^{[2]}}{F_{d}^{[2]}},\label{25}\\
&w[2]_{d}=-\dfrac{\mu u_{0}}{e_{0}}+4i\dfrac{\partial}{\partial t}\dfrac{H_{d}^{[2]}}{F_{d}^{[2]}},\label{26}
\end{align}
and
\begin{equation}\label{27}
u[2]_{d}=u_{0}+\dfrac{4i}{\mu}\left(\dfrac{\partial^2}{\partial x\partial t}
\dfrac{G_{d}^{[2]}}{F_{d}^{[2]}}-
e_{0}\dfrac{\partial }{\partial t}\dfrac{H_{d}^{[2]}}{F_{d}^{[2]}}-\dfrac{\mu u_{0}}{e_{0}}\dfrac{G_{d}^{[2]}}{F_{d}^{[2]}}
+4i\dfrac{G_{d}^{[2]}}{F_{d}^{[2]}}\dfrac{\partial }
{\partial t}\dfrac{H_{d}^{[2]}}{F_{d}^{[2]}}\right),
\end{equation}
where the explicit expressions of the mixed functions containing rational polynomials and
trigonometric functions are given in appendix A.

\begin{figure}[!h]
\centering
\renewcommand{\figurename}{{\bf Fig.}}
{\includegraphics[height=10cm,width=16cm]{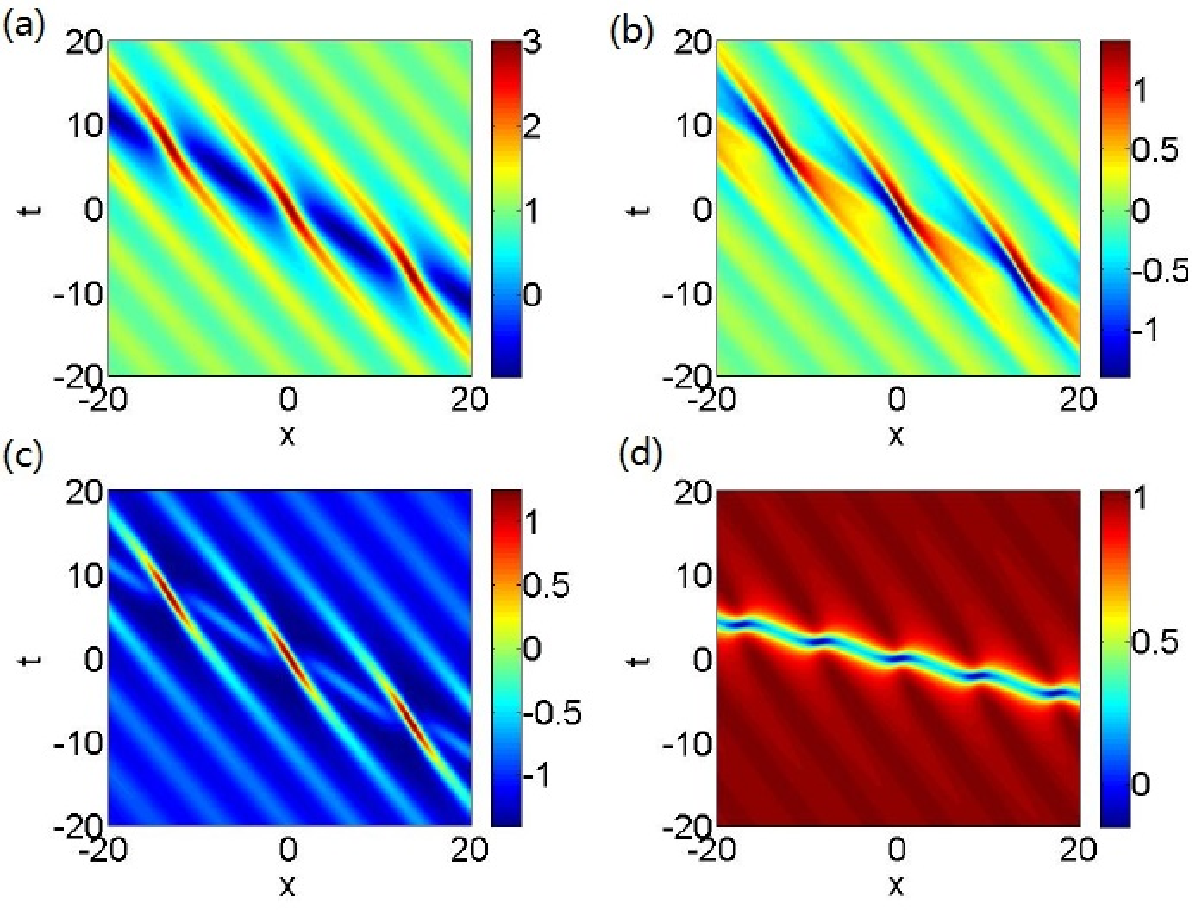}}
\caption{(a)-(d) The degenerate periodic solutions (\ref{24})-(\ref{26}) for the components $E$, $v$, $w$
with $\mu=1$ and (\ref{27}) for the component $u$ with $\mu=0.2$. The other parameters are $e_{0}=1,u_{0}=1$.}
\label{fig:3}
\end{figure}

For $N=1$, the degenerate periodic solutions in the above formulas are reduced to the first-order periodic solutions (\ref{10})-(\ref{13}) given in the above section. For $N=2$, the simplest nontrivial degenerate
periodic solutions can be shown, see Fig. \ref{fig:3}. The patterns of these solutions
for the components $E$, $v$ and $w$ consist of a single periodic- peaks on a periodic wave background,
and the maximum amplitudes of the peaks are 3, 1.40 and 1.24, respectively.
While for the component $u$, it is exhibited that a singe periodic- dips on a periodic wave background
structure exists, and the minimum amplitude of the dips is -0.15.

\section{Rational solutions}

In this section, let us take a further look at the general periodic solutions by employing the
limit approach of $\epsilon_{1}\rightarrow0$.  We now adjust the expression $A$ in (\ref{05}) as
$$
A=e_{0}\epsilon\sqrt{1+\epsilon^2}\left[x+\dfrac{\mu u_{0}}{e_{0}(4e_{0}^2\epsilon^4
+4e_{0}^2\epsilon^2+e_{0}^2+\mu^2)}t+\sum_{i=1}^{N-1}s_{i}\epsilon^{2i}   \right],
$$
here $s_{i}$ are ($N-1$) new introduced complex free parameters. Accordingly,
the Taylor series (\ref{18}) and (\ref{19}) can be rewritten as
\begin{equation}\label{28}
\psi(\epsilon)=\sum_{i=0}^{N-1}\widehat{\psi_{1}^{[i]}}\epsilon^{2i}
+\mathcal{O}\left(\epsilon^{2N}\right),
\varphi(\epsilon)=\sum_{i=0}^{N-1}\widehat{\varphi_{1}^{[i]}}\epsilon^{2i}
+\mathcal{O}\left(\epsilon^{2N}\right),
\end{equation}
and
\begin{equation}\label{29}
\dfrac{\psi(\epsilon^{*})\psi(\epsilon)+\varphi(\epsilon^{*})\varphi(\epsilon)}{ie_{0}(1+\epsilon^2+\epsilon^{*2})}
=\sum_{i,j=1}^{N}Q^{[i,j]}\epsilon^{2(j-1)}\epsilon^{*2(i-1)}+
\mathcal{O}\left((\epsilon\epsilon^{*})^{2N}\right),
\end{equation}
where
$$
\widehat{\psi_{1}^{[i]}}=\lim\limits_{\epsilon\rightarrow0}
\frac{\displaystyle1}{\displaystyle 2i!}\frac{\displaystyle\partial^{2i}\psi_{1}}{\displaystyle\partial \epsilon^{2i}},\widehat{\varphi_{1}^{[i]}}=\lim\limits_{\epsilon\rightarrow0}
\frac{\displaystyle1}{\displaystyle 2i!}\frac{\displaystyle\partial^{2i}\varphi_{1}}{\displaystyle\partial \epsilon^{2i}},
$$
and
$$
Q^{[i,j]}=\dfrac{1}{2(i-1)!2(j-1)!}\dfrac{\partial ^{2(i+j-2)}}{\partial \epsilon^{2(j-1)}\partial \epsilon^{*2(i-1)}}
\dfrac{\psi(\epsilon^{*})\psi(\epsilon)+\varphi(\epsilon^{*})\varphi(\epsilon)}{ie_{0}(1+\epsilon^2+\epsilon^{*2})}
\bigg|_{\epsilon,\epsilon^{*}\rightarrow0}.
$$

At present, we can put forward the $N$th-order rational solution as
\begin{align}
&E[N]=e_{0}-4i\dfrac{\det(Q_{1})}{\det(Q)},\label{en}\\
&v[N]=4i\dfrac{\partial}{\partial t}\dfrac{\det(Q_{1})}{\det(Q)},\label{vn}\\
&w[N]=-\dfrac{\mu u_{0}}{e_{0}}+4i\dfrac{\partial}{\partial t}\dfrac{\det(Q_{2})}{\det(Q)},\label{wn}
\end{align}
and
\begin{equation}\label{un}
u[N]=u_{0}+\dfrac{4i}{\mu}\left[\dfrac{\partial^2}{\partial x\partial t}\dfrac{\det(Q_{1})}{\det(Q)}-
e_{0}\dfrac{\partial }{\partial t}\dfrac{\det(Q_{2})}{\det(Q)}
-\dfrac{\mu u_{0}}{e_{0}}\dfrac{\det(Q_{1})}{\det(Q)}+4i\dfrac{\det(Q_{1})}{\det(Q)}\dfrac{\partial }{\partial t}\dfrac{\det(Q_{2})}{\det(Q)}\right],
\end{equation}
where
$$\begin{array}{l}
Q=\left(
    \begin{array}{cccc}
      Q^{[11]} & Q^{[12]} & \cdots & Q^{[1N]} \\
       Q^{[21]} & Q^{[22]} & \cdots & Q^{[2N]} \\
      \vdots & \vdots & \ddots & \vdots \\
       Q^{[N1]} & Q^{[N2]} & \cdots & Q^{[NN]} \\
    \end{array}
  \right),\\
Q_{1}=\left(
    \begin{array}{ccccc}
      Q^{[11]} & Q^{[12]} & \cdots & Q^{[1N]} & \widehat{\varphi_{1}^{[0]}}\\
       Q^{[21]} & Q^{[22]} & \cdots & Q^{[2N]}  & \widehat{\varphi_{1}^{[1]}}\\
      \vdots & \vdots & \ddots & \vdots \vdots\\
       Q^{[N1]} & Q^{[N2]} & \cdots & Q^{[NN]} & \widehat{\varphi_{1}^{[N-1]}}\\
       \widehat{\psi_{1}^{[0]}} &\widehat{\psi_{1}^{[1]}} &\cdots &\widehat{\psi_{1}^{[N-1]}}& 0
    \end{array}
  \right),
\end{array}
$$
$$\begin{array}{l}
Q_{2}=\left(
    \begin{array}{ccccc}
      Q^{[11]} & Q^{[12]} & \cdots & Q^{[1N]} & \widehat{\psi_{1}^{[0]}}\\
        Q^{[21]} & Q^{[22]} & \cdots & Q^{[2N]}  & \widehat{\psi_{1}^{[1]}}\\
      \vdots & \vdots & \ddots & \vdots \vdots\\
       Q^{[N1]} & Q^{[N2]} & \cdots & Q^{[NN]} & \widehat{\psi_{1}^{[N-1]}}\\
       \widehat{\psi_{1}^{[0]}} &\widehat{\psi_{1}^{[1]}} &\cdots &\widehat{\psi_{1}^{[N-1]}}& 0
    \end{array}
  \right).
\end{array}
$$

With these formulas for $N=1$, we can get the explicit first-order rational solutions, viz.
\begin{align}
&E[1]_{r}=-e_{0}+\dfrac{4e_{0}(e_{0}^4+2e_{0}^2\mu^2+\mu^4)}{F_{r}^{[1]}},\label{34}\\
&v[1]_{r}=\dfrac{8 e_{0}\mu u_{0}(e_{0}^2+\mu^2)^2(e_{0}^3x+e_{0}\mu^2x+\mu u_{0} t)}{F_{r}^{[1]2}},\label{35}\\
&w[1]_{r}=-\dfrac{\mu u_{0}}{e_{0}}-\dfrac{4 e_{0}\mu u_{0}(e_{0}^2+\mu^2)H_{r}^{[1]}}{F_{r}^{[1]2}},\label{36}\\
&u[1]_{r}=-u_{0}+\dfrac{4 \mu^2 u_{0}(e_{0}^2+\mu^2)}{F_{r}^{[1]}},\label{37}
\end{align}
where
$$\begin{array}{l}
H_{r}^{[1]}=(e_{0}^3x+e_{0}\mu^2x+\mu u_{0} t+e_{0}^2+\mu^2)(e_{0}^3x+e_{0}\mu^2x+\mu u_{0}t-e_{0}^2-\mu^2),\\
F_{r}^{[1]}=e_{0}^6x^2+2e_{0}^4\mu^2x^2+e_{0}^2\mu^4x^2+2e_{0}^3\mu u_{0}x t+2e_{0}\mu^3tu_{0}x
+\mu^2t^2u_{0}^2+e_{0}^4+2e_{0}^2\mu^2+\mu^4.
\end{array}
$$

\begin{figure}[!h]
\centering
\renewcommand{\figurename}{{\bf Fig.}}
{\includegraphics[height=10cm,width=16cm]{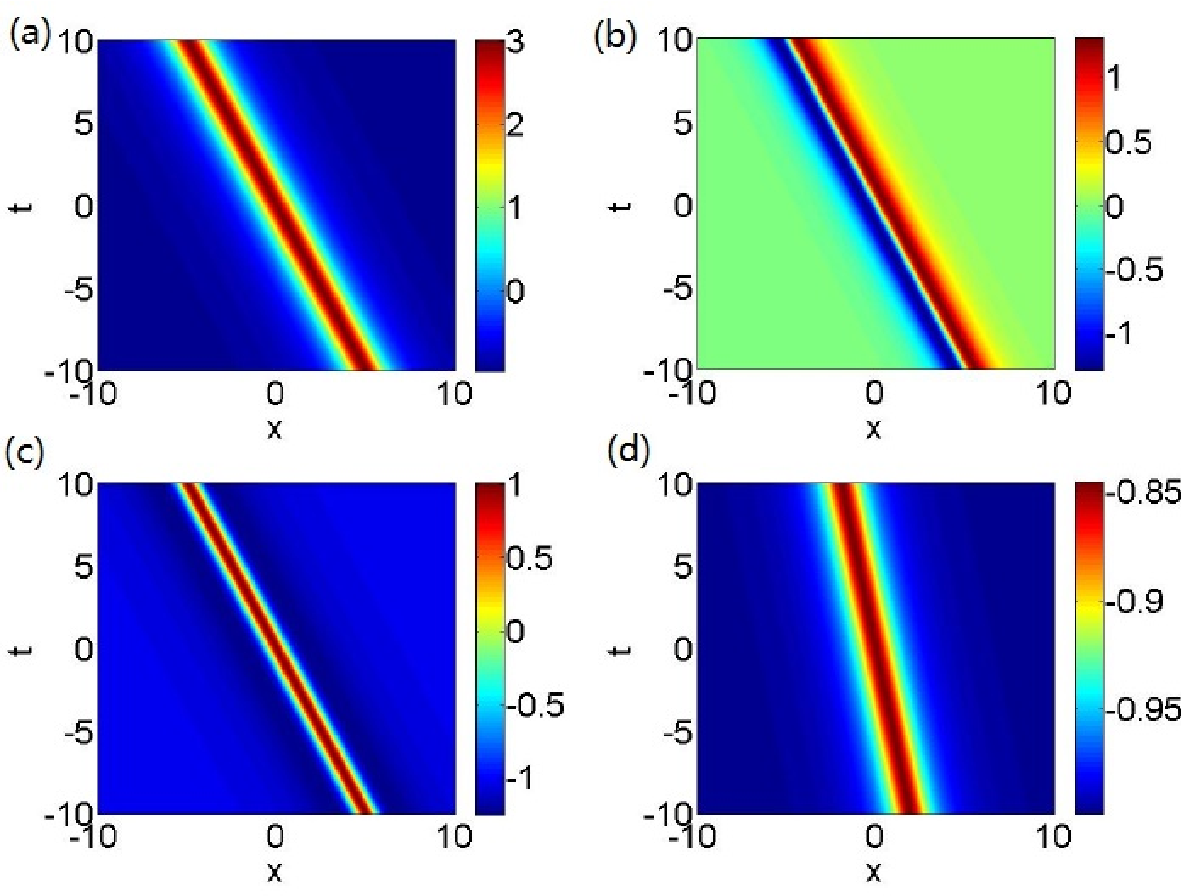}}
\caption{(a)-(d) The first-order rational solutions (\ref{34})-(\ref{36}) for the components $E$, $v$, $w$
with $\mu=1$ and (\ref{37}) for the component $u$ with $\mu=0.2$.
The other parameters are $e_{0}=1,u_{0}=1$.}
\label{fig:4}
\end{figure}

The above solutions correspond to the Peregrine soliton of the NLS equation \cite{13}. Nevertheless, unlike the Peregrine soliton that is doubly localized, these solutions look more like the solitons on a
finite constant background. For the component $E$, there is one ridge with maximum amplitude $3e_{0}$ on the
temporal-spatial distribution, see Fig. \ref{fig:4}(a). The critical line is given by
\begin{equation}\label{tr}
t=-\dfrac{e_{0}(e_{0}^2+\mu^2)}{\mu u_{0}}x.
\end{equation}
For the component $v$,
it is displayed in Fig. \ref{fig:4}(b) that, there are one ridge with maximum amplitude and
one valley with minimum  amplitude
on the coordinate plane. The maximum and minimum values of $v[1]$ are
$\pm3\sqrt{3}e_{0}\mu u_{0}/2(e_{0}^2+\mu^2)$, and occur at
$$
t=-\dfrac{e_{0}(e_{0}^2+\mu^2)}{\mu u_{0}}x\pm\dfrac{\sqrt{3}(e_{0}^2+\mu^2)}{3\mu u_{0}}.
$$
Moreover, $w[1]$ is shown in Fig. \ref{fig:4}(c), and there exist one ridge with maximum amplitude
and two valleys with minimum amplitude  on the temporal-spatial plane.
The maximum value of $w[1]$ is $(3e_{0}^2-\mu^2)\mu u_{0}/e_{0}(e_{0}^2+\mu^2)$ and the
critical line is defined by Eq. (\ref{tr}). While the minimum value of $w[1]$ is
$-(3e_{0}^2+2\mu^2)\mu u_{0}/2e_{0}(e_{0}^2+\mu^2)$ and is reached at
$$
t=-\dfrac{e_{0}(e_{0}^2+\mu^2)}{\mu u_{0}}x\pm\dfrac{\sqrt{3}(e_{0}^2+\mu^2)}{\mu u_{0}}.
$$
In succession, the maximum amplitude of $u[1]$ is $-(e_{0}^2-3\mu^2)u_{0}/(e_{0}^2+\mu^2)$
and the critical line is defined by Eq. (\ref{tr}), see Fig. \ref{fig:4}(d).

Additionally, it can be checked that
$$\int_{-\infty}^{\infty}(E[1]_{r}-E[1]_{r0})^2{\rm d}x=8\pi {\rm sgn}(e_{0})e_{0},$$
$$\int_{-\infty}^{\infty}(v[1]_{r}-v[1]_{r0})^2{\rm d}x=\int_{-\infty}^{\infty}(w[1]_{r}-w[1]_{r0})^2{\rm d}x
=\dfrac{4\pi \mu^2u_{0}^2{\rm sgn}(e_{0})e_{0}}{(e_{0}^2+\mu^2)^2},$$
$$\int_{-\infty}^{\infty}(u[1]_{r}-u[1]_{r0})^2{\rm d}x=\dfrac{8\pi \mu^4u_{0}^2{\rm sgn}(e_{0})}{e_{0}(e_{0}^2+\mu^2)^2},$$
where $E[1]_{r0},v[1]_{r0},w[1]_{r0},w[1]_{r0}=\displaystyle\lim_{x\rightarrow \pm\infty}E[1]_{r},v[1]_{r},
w[1]_{r},u[1]_{r}$,
which indicate that
energies of the Peregrine pulses of the first-order rational solutions keep a constant.

Afterwards, by applying $N=2$ in the formulas (\ref{en})-(\ref{un}), we can present
the second-order rational solutions
\begin{align}
&E[2]_{r}=e_{0}-4i\dfrac{G_{r}^{[2]}}{F_{r}^{[2]}},\label{38}\\
&v[2]_{r}=4i\dfrac{\partial}{\partial t}\dfrac{G_{r}^{[2]}}{F_{r}^{[2]}},\label{39}\\
&w[2]_{r}=-\dfrac{\mu u_{0}}{e_{0}}+4i\dfrac{\partial}{\partial t}\dfrac{H_{r}^{[2]}}{F_{r}^{[2]}},\label{40}
\end{align}
and
\begin{equation}\label{41}
u[2]_{r}=u_{0}+\dfrac{4i}{\mu}\left(\dfrac{\partial^2}{\partial x\partial t}
\dfrac{G_{r}^{[2]}}{F_{r}^{[2]}}-
e_{0}\dfrac{\partial }{\partial t}\dfrac{H_{r}^{[2]}}{F_{r}^{[2]}}-\dfrac{\mu u_{0}}{e_{0}}\dfrac{G_{r}^{[2]}}{F_{r}^{[2]}}
+4i\dfrac{G_{r}^{[2]}}{F_{r}^{[2]}}\dfrac{\partial }
{\partial t}\dfrac{H_{r}^{[2]}}{F_{r}^{[2]}}\right),
\end{equation}
where the polynomials are explicitly provided in appendix B.

\begin{figure}[!h]
\centering
\renewcommand{\figurename}{{\bf Fig.}}
{\includegraphics[height=10cm,width=16cm]{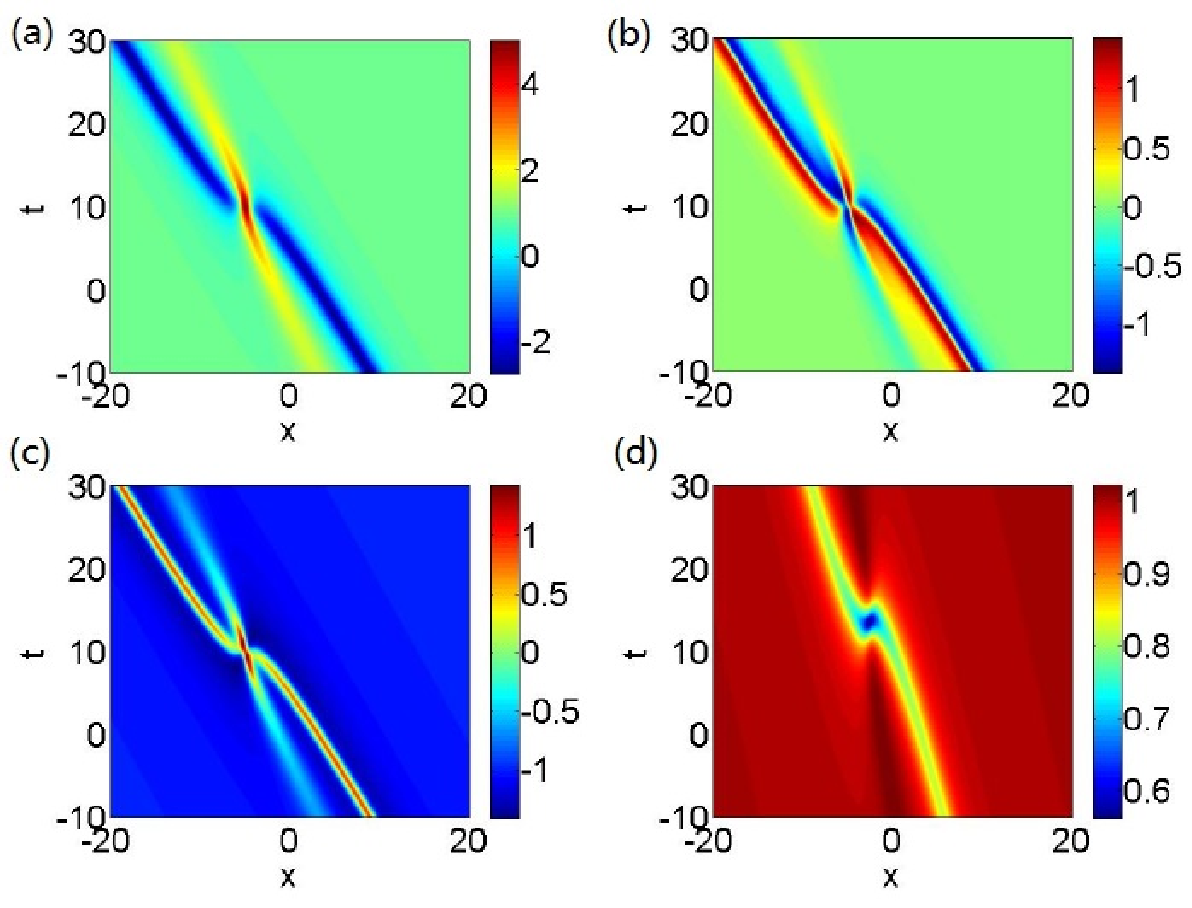}}
\caption{(a)-(d) The second-order rational solutions (\ref{38})-(\ref{40}) for the components $E$, $v$, $w$
with $\mu=1$ and (\ref{41}) for the component $u$ with $\mu=0.2$.
The other parameters are $e_{0}=1,u_{0}=1,s_{1}=10$.}
\label{fig:6}
\end{figure}

Shown in Fig. \ref{fig:5} are the second-order rational solutions (\ref{38})-(\ref{41}).
It is exhibited that some typical nonlinear wave patterns can emerge and they are seemingly
a result of the collisions of a fixed number of dark and bright solitons.
For the component $E$, we can see
the single-peak structure in Fig. \ref{fig:5}(a). The maximum amplitude of $E[2]$ is 5
and is localized at $(0,0)$, which coincides with the second-order fundamental
rogue wave in the NLS equation \cite{14,28}. For the component $v$, it is interestingly
found that the double-peak-double-dip structure can arise, see Fig. \ref{fig:5}(b).
We calculate that the maximum amplitude of $v[2]$ is 1.41 and occurs at $(0.05,1.18)$ and $(0.94,-1.40)$,
and the minimum amplitude of it is -1.41 and arrives at $(-0.05,-1.18)$ and $(-0.94,1.40)$.
Furthermore, the nonlinear wave in Fig. \ref{fig:5}(c) for the component $w$ is the double-peak structure,
which has the maximum amplitude of 1.41 and reaches at the two points of $(0.27,-0.91)$ and $(-0.27,0.91)$.
More interestingly, it is worthwhile to emphasize that, for the component $u$ in Fig. \ref{fig:5}(d),
the single-dip structure can be exhibited, which is
quite different from the single-peak case for the component $E$ and the relevant
structure in the mKdV equation \cite{22}. The minimum amplitude of $u[2]$ is 0.56 and
is acquired at the original point.  At the same time, we can see that in Fig. \ref{fig:6},
the nonzero free parameter $s_{1}$ can produce an important shift effect
on the peaks or dips along the wave trough in the second-order rational solutions.
But, it does not affect the maximum or minimum amplitudes of these solutions.

\begin{figure}[!h]
\centering
\renewcommand{\figurename}{{\bf Fig.}}
{\includegraphics[height=5cm,width=16cm]{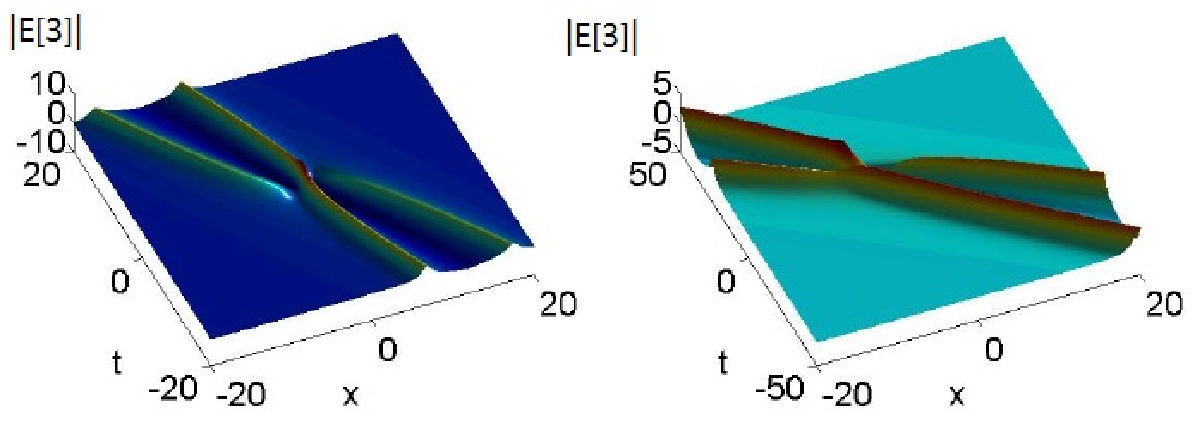}}
\begin{center}
\hskip 1cm $(\rm{a})$ \hskip 6cm $(\rm{b})$
\end{center}
\caption{(a),(b) The third-order rational solutions for the component $E$ with
$s_{1}=0$ and $s_{1}=10$. The other parameters are
$e_{0}=1,\mu=1,u_{0}=1,s_{2}=0$.}
\label{fig:7}
\end{figure}

\begin{figure}[!h]
\centering
\renewcommand{\figurename}{{\bf Fig.}}
{\includegraphics[height=5cm,width=16cm]{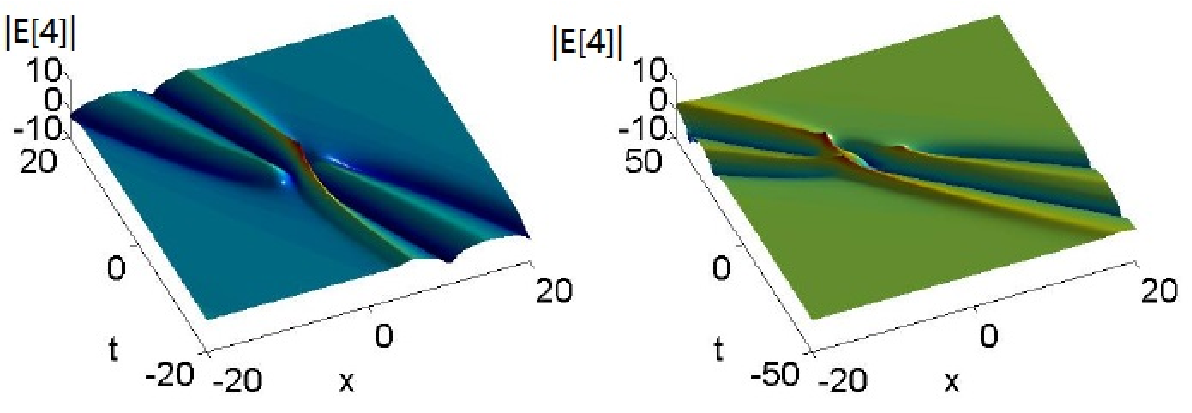}}
\begin{center}
\hskip 1cm $(\rm{a})$ \hskip 6cm $(\rm{b})$
\end{center}
\caption{(a),(b) The fourth-order rational solutions for the component $E$ with
$s_{1}=0$ and $s_{1}=10$. The other parameters are
$e_{0}=1,\mu=1,u_{0}=1,s_{2}=s_{3}=0$.}
\label{fig:8}
\end{figure}

For the higher-order cases, we only show the evolution plots of the third- and fourth-order
rational solutions for the component $E$, and omit writing down the corresponding cumbersome expressions,
see Figs. \ref{fig:7} and \ref{fig:8}. We exhibit that
the third-order rational solutions can be viewed as the collisions of a dark and two bright solitons,
and the fourth-order rational solutions are likely the result of the collisions
of two dark and two bright solitons.
It is computed that when setting all of the free parameters $s_{i}$ be zero, then
the maximum amplitudes of the third- and fourth-order rational solutions
are 7 and 9, respectively, which are
the same as that of the third- and fourth-order rogue waves in the NLS equation \cite{14,28}.
While when taking one of the free parameters be nonzero such as $s_{1}\neq0$, then the highest peaks
in these higher-order rational solutions can also have a shift along the depressed wave trough.
In this circumstance, unlike the second-order rational solutions, the maximum amplitudes
of the higher-order solutions are changed due to the interactions among the multiple
solitons. The maximum amplitudes of the highest peaks in Figs. \ref{fig:7}(b) and \ref{fig:8}(b)
become 4.77 and 5.96, respectively.

\section{Conclusion}

In summary, we proposed the general $N$th-order periodic, $N$th-order degenerate periodic and
$N$th-order rational solutions with the compact determinant representations
for the RMB equations, which serve as the fundamental model in
nonlinear optics associated with self-induced transparency.
The explicit first- and second-order periodic and rational solutions are presented.
Some interesting nonlinear wave patterns described by the
second-order periodic solution, the simplest degenerate periodic solutions, and especially the
higher-order rational solutions are shown.
Also, it is notable to remark that the limit approach using in this paper
can be directly applied to the mKdV equation \cite{22,He}, the variable-coefficient mKdV equation \cite{vmkdv}
and other mKdV-type equations.  We hope
our results given in this paper may be helpful to interpret
the intricate rogue wave phenomena in nonlinear optics governed by the RMB equations.

\section*{Appendix A: mixed functions in Eqs. (\ref{24})-(\ref{27})}

$$
\begin{array}{l}
F_{d}^{[2]}=-\dfrac{256}{27e_{0}^2(e_{0}^2+4\mu^4)^4}[-2352e_{0}^4\mu^2u_{0}^2t^2-2688e_{0}^2\mu^4u_{0}^2t^2
-768\mu^6u_{0}^2t^2-168e_{0}^7\mu~u_{0}tx\\
~~~~~-1440e_{0}^5\mu^3u_{0}tx-3456e_{0}^3\mu^5u_{0}tx-1536e_{0}\mu^7u_{0}tx-3e_{0}^{10}x^2
-48e_{0}^8\mu^2x^2-288e_{0}^6\mu^4x^2\\
~~~~~-768e_{0}^4\mu^6x^2-768e_{0}^2\mu^8x^2+16(e_{0}^2+4\mu^2)^4\cos(\omega)^4+80(e_{0}^2+4\mu^2)^4\cos(\omega)^2\\
~~~~~+16\sqrt{3}(28e_{0}^2\mu u_{0}t+16\mu^3u_{0}t+e_{0}^5x+8e_{0}^3\mu^2x +16e_{0}\mu^4x)(e_{0}^2+4\mu^2)^2\sin(\omega)\cos(\omega)\\
~~~~~-108e_{0}^8-1728e_{0}^6\mu^2-10368e_{0}^4\mu^4
-27648e_{0}^2\mu^6-27648\mu^8],\\
\end{array}
$$

$$
\begin{array}{l}
F_{d}^{[2]}=\dfrac{512i}{9e_{0}(e_{0}^2+4\mu^4)^4}[10(e_{0}^2+4\mu^2)^2\cos(\omega)^2+\sqrt{3}(28e_{0}^2\mu u_{0}t+16\mu^3u_{0}t+e_{0}^5x+8e_{0}^3\mu^2x\\
~~~~~+16e_{0}\mu^4x)\sin(\omega)\cos(\omega)-9(e_{0}^2+4\mu^2)^2],\\
H_{d}^{[2]}=-\dfrac{256}{27e_{0}^2(e_{0}^2+4\mu^4)^4}[2352e_{0}^4\mu^2u_{0}^2t^2
+2688e_{0}^2\mu^4u_{0}^2t^2+768\mu^6u_{0}^2t^2+1536 e_{0} \mu^7 u_{0} t x\\
~~~~~+3456 e_{0}^3 \mu^5 u_{0} t x+168 e_{0}^7 \mu u_{0} t x+1440 e_{0}^5 \mu^3 u_{0} t x+288 e_{0}^6 \mu^4 x^2+768 e_{0}^4 \mu^6 x^2+768 e_{0}^2 \mu^8 x^2\\
~~~~~+3 e_{0}^{10} x^2+48 e_{0}^8 \mu^2 x^2+4608 \mu^7 u_{0} t +504 e_{0}^6 \mu u_{0} t+4320 e_{0}^4 \mu^3 u_{0} t+10368 e_{0}^2 \mu^5 u_{0} t\\
~~~~~+288 e_{0}^7 \mu^2 x+1728 e_{0}^5 \mu^4 x+4608 e_{0}^3 \mu^6 x+4608 e_{0} \mu^8 x+18 e_{0}^9 x+10368 e_{0}^4 \mu^4+27648 e_{0}^2 \mu^6\\
~~~~~+16 \sqrt{3} (e_{0}^2+4 \mu^2)^4 \sin(\omega) \cos(\omega)^3-8 (e_{0}^2+4 \mu^2)^2 (84 e_{0}^2 \mu u_{0} t+48 \mu^3 u_{0} t+3 e_{0}^5 x\\
~~~~~+24 e_{0}^3 \mu^2 x+48 e_{0} \mu^4 x+10 e_{0}^4+80 e_{0}^2 \mu^2+160 \mu^4) \cos(\omega)^2-16 (e_{0}^2+4 \mu^2)^4 \cos(\omega)^4\\
~~~~~+1728 e_{0}^6 \mu^2+27648 \mu^8+108 e_{0}^8-8 \sqrt{3} (56 e_{0}^2 \mu u_{0} t+32 \mu^3 u_{0} t+2 e_{0}^5 x+16 e_{0}^3 \mu^2 x\\
~~~~~+32 e_{0} \mu^4 x-3 e_{0}^4-24 e_{0}^2 \mu^2-48 \mu^4) (e_{0}^2+4 \mu^2)^2 \sin(\omega) \cos(\omega)],
\end{array}
$$
in which $\omega=\sqrt{3}(e_{0}^3 x+4 e_{0}\mu^2 x+4\mu u_{0}t)/4(e_{0}^2+4\mu^2).$

\section*{Appendix B: polynomials in Eqs. (\ref{38})-(\ref{41}) }

$$
\begin{array}{l}
F_{r}^{[2]}=-\dfrac{4}{9e_{0}^2(e_{0}^2+\mu^2)^6}[\mu^6 u_{0}^6 t^6+6 e_{0} \mu^5 u_{0}^5 (e_{0}^2+\mu^2) t^5 x+e_{0}^6 (e_{0}^2+\mu^2)^6 x^6+(15 e_{0}^2 \mu^4 u_{0}^4 (e_{0}^2+\mu^2)^2 x^2\\
~~~~~+3 \mu^4 u_{0}^4 (9 e_{0}^2+\mu^2) (e_{0}^2+\mu^2)) t^4+3 e_{0}^4 (e_{0}^2+\mu^2)^6 x^4+(20 e_{0}^3 \mu^3 u_{0}^3 (e_{0}^2+\mu^2)^3 x^3\\
~~~~~+12 e_{0} \mu^3 u_{0}^3 (7 e_{0}^2+\mu^2) (e_{0}^2+\mu^2)^2 x-6 e_{0} \mu^3 s_{1} u_{0}^3 (e_{0}^2+\mu^2)^3) t^3-6 e_{0}^4 s_{1} (e_{0}^2+\mu^2)^6 x^3\\
~~~~~+(15 e_{0}^4 \mu^2 u_{0}^2 (e_{0}^2+\mu^2)^4 x^4+18 e_{0}^2 \mu^2 u_{0}^2 (5 e_{0}^2+\mu^2) (e_{0}^2+\mu^2)^3 x^2-18 e_{0}^2 \mu^2 s_{1} u_{0}^2 (e_{0}^2\\
~~~~~+\mu^2)^4 x+9 \mu^2 u_{0}^2 (11 e_{0}^4-2 e_{0}^2 \mu^2+3 \mu^4) (e_{0}^2+\mu^2)^2) t^2+27 e_{0}^2 (e_{0}^2+\mu^2)^6 x^2\\
~~~~~+(6 e_{0}^5 \mu u_{0} (e_{0}^2+\mu^2)^5 x^5+12 e_{0}^3 \mu u_{0} (3 e_{0}^2+\mu^2) (e_{0}^2+\mu^2)^4 x^3
-18 e_{0}^3 \mu s_{1} u_{0} (e_{0}^2\\
~~~~~+\mu^2)^5 x^2-18 e_{0}\mu u_{0} (e_{0}^2-3 \mu^2) (e_{0}^2+\mu^2)^4 x-18 e_{0} \mu s_{1} u_{0} (3 e_{0}^2
-\mu^2) (e_{0}^2+\mu^2)^4) t\\
~~~~~+18 e_{0}^2 s_{1} (e_{0}^2+\mu^2)^6 x+(9 (e_{0}^2 s_{1}^2+1)) (e_{0}^2+\mu^2)^6],\\

G_{r}^{[2]}=\dfrac{4i}{3e_{0}^2(e_{0}^2+\mu^2)^6}
[\mu^4 u_{0}^4 t^4+4 e_{0} \mu^3 u_{0}^3 (e_{0}^2+\mu^2) t^3 x+e_{0}^4 (e_{0}^2+\mu^2)^4 x^4
+(6 \mu^2 u_{0}^2 e_{0}^2 (e_{0}^2+\mu^2)^2 x^2\\
~~~~~-6 \mu^2 u_{0}^2 (3 e_{0}^2-\mu^2) (e_{0}^2+\mu^2)) t^2
+6 e_{0}^2 (e_{0}^2+\mu^2)^4 x^2+(4 e_{0}^3 \mu u_{0} (e_{0}^2+\mu^2)^3 x^3\\
~~~~~+12 u_{0} \mu e_{0} (e_{0}^2
-\mu^2) (e_{0}^2+\mu^2)^2 x+6 e_{0} \mu s_{1} u_{0} (e_{0}^2+\mu^2)^3) t+6 e_{0}^2 s_{1} (e_{0}^2
+\mu^2)^4 x\\~~~~~-3 (e_{0}^2+\mu^2)^4],\\

H_{r}^{[2]}=\dfrac{4i}{9e_{0}^2(e_{0}^2+\mu^2)^6}[\mu^6 u_{0}^6 t^6+e_{0}^6 (e_{0}^2
+\mu^2)^6 x^6+(6 e_{0} \mu^5 u_{0}^5 (e_{0}^2+\mu^2) x+3 \mu^5 u_{0}^5 (e_{0}^2+\mu^2)) t^5\\
~~~~~+3 e_{0}^5 (e_{0}^2+\mu^2)^6 x^5+(15 e_{0}^2 \mu^4 u_{0}^4 (e_{0}^2
+\mu^2)^2 x^2+15 e_{0} \mu^4 u_{0}^4 (e_{0}^2+\mu^2)^2 x\\
~~~~~+3 \mu^4 u_{0}^4 (9 e_{0}^2+\mu^2) (e_{0}^2+\mu^2)) t^4+3 e_{0}^4 (e_{0}^2+\mu^2)^6 x^4+(20 e_{0}^3 \mu^3 u_{0}^3 (e_{0}^2\\
~~~~~+\mu^2)^3 x^3+30 e_{0}^2 \mu^3 u_{0}^3 (e_{0}^2+\mu^2)^3 x^2+12 e_{0} \mu^3 u_{0}^3 (7 e_{0}^2+\mu^2) (e_{0}^2+\mu^2)^2 x\\
~~~~~-6 \mu^3 u_{0}^3 (e_{0}^2+\mu^2)^2 (e_{0}^3 s_{1}+e_{0} \mu^2 s_{1}
-7 e_{0}^2-\mu^2)) t^3-6 e_{0}^3 (e_{0}^2+\mu^2)^6 (e_{0} s_{1}-1) x^3\\
~~~~~+(15 e_{0}^4 \mu^2 u_{0}^2 (e_{0}^2+\mu^2)^4 x^4+30 e_{0}^3 \mu^2 u_{0}^2 (e_{0}^2+\mu^2)^4 x^3+18 e_{0}^2 \mu^2 u_{0}^2 (5 e_{0}^2\\
~~~~~+\mu^2) (e_{0}^2+\mu^2)^3 x^2-18 e_{0} \mu^2 u_{0}^2 (e_{0}^2+\mu^2)^3 (e_{0}^3 s_{1}+e_{0} \mu^2 s_{1}-5 e_{0}^2-\mu^2) x\\
~~~~~-9 \mu^2 u_{0}^2 (e_{0}^2+\mu^2)^2 (e_{0}^5 s_{1}+2 e_{0}^3 \mu^2 s_{1}+e_{0} \mu^4 s_{1}
-11 e_{0}^4+2 e_{0}^2 \mu^2\\
~~~~~-3 \mu^4)) t^2-9 e_{0}^2 (e_{0}^2+\mu^2)^6 (e_{0} s_{1}-3) x^2
+(6 e_{0}^5 \mu u_{0} (e_{0}^2+\mu^2)^5 x^5+15 e_{0}^4 \mu u_{0} (e_{0}^2+\mu^2)^5 x^4\\
~~~~~+12 e_{0}^3 \mu u_{0} (3 e_{0}^2+\mu^2) (e_{0}^2+\mu^2)^4 x^3-18 e_{0}^2 \mu u_{0} (e_{0}^2
+\mu^2)^4 (e_{0}^3 s_{1}\\
~~~~~+e_{0} \mu^2 s_{1}-3 e_{0}^2-\mu^2) x^2-18 e_{0}\mu u_{0} (e_{0}^2
+\mu^2)^4 (e_{0}^3 s_{1}+e_{0} \mu^2 s_{1}+e_{0}^2-3 \mu^2) x\\
~~~~~-9 \mu u_{0} (e_{0}^2+\mu^2)^4 (6 e_{0}^3 s_{1}
-2 e_{0} \mu^2 s_{1}+e_{0}^2-3 \mu^2)) t+9 e_{0} (e_{0}^2+\mu^2)^6 (2 e_{0} s_{1}+3) x
\\~~~~~+9 (e_{0}^2 s_{1}^2+e_{0} s_{1}+1) (e_{0}^2+\mu^2)^6].
\end{array}
$$

\section*{Acknowledgment}
One of the authors X. Wang would like to thank Prof. Y. Chen for his valuable
suggestions and enthusiastic guidances.
This work is supported by National Natural Science Foundation of China (11331008),
China Postdoctoral Science Foundation funded sixtieth batches
(No. 2016M602252).

\newpage

\begin{figure}[!h]
\centering
\renewcommand{\figurename}{{\bf Fig.}}
{\includegraphics[height=10cm,width=16cm]{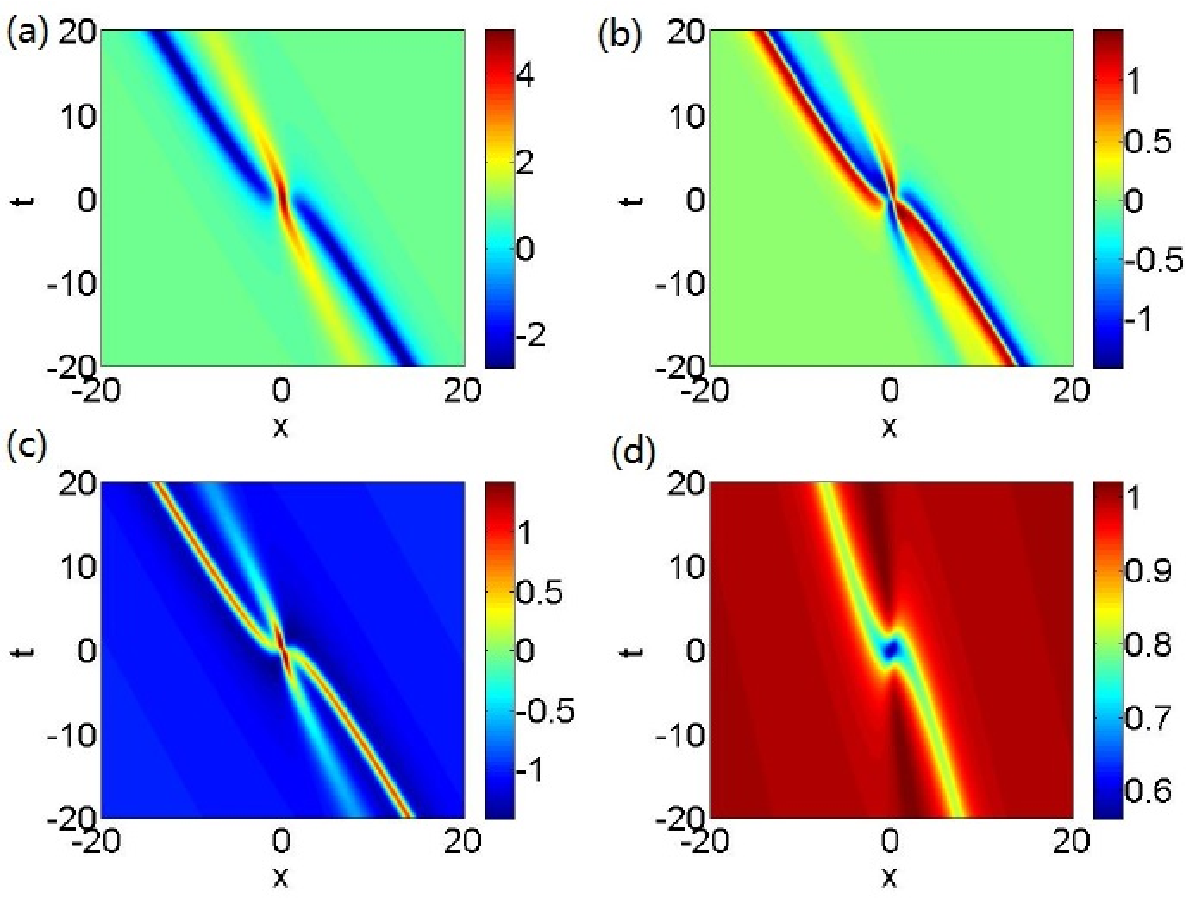}}
\caption{(a)-(d) The second-order rational solutions (\ref{38})-(\ref{40}) for the components $E$, $v$, $w$
with $\mu=1$ and (\ref{41}) for the component $u$ with $\mu=0.2$.
The other parameters are $e_{0}=1,u_{0}=1,s_{1}=0$.}
\label{fig:5}
\end{figure}

\end{document}